\documentclass[conference]{IEEEtran}
\IEEEoverridecommandlockouts
\usepackage{cite}
\usepackage{amsmath,amssymb,amsfonts}
\usepackage{algorithmic}
\usepackage{graphicx}
\usepackage{textcomp}
\usepackage{xcolor}
\usepackage{pifont}
\usepackage{hyperref}
\usepackage{cleveref}
\usepackage{amsmath,amsfonts}
\usepackage{multirow}
\usepackage{wrapfig}
\usepackage[ruled,lined,linesnumbered,vlined,algo2e]{algorithm2e}
\usepackage{graphicx}
\usepackage{textcomp}
\usepackage{xcolor}
\usepackage{framed}
\usepackage{listings}
\usepackage{bbding} 
\usepackage{color}
\usepackage{enumitem}
\usepackage{threeparttable}
\usepackage{xspace}
\usepackage{indentfirst}
\usepackage{booktabs}
\usepackage{subfig}
\usepackage{balance}
\usepackage{colortbl}
\usepackage{comment}
\usepackage{url}
\usepackage{cite} 
\usepackage[justification=centering]{caption}
\usepackage[ruled,lined,linesnumbered,vlined,algo2e]{algorithm2e}
\def\BibTeX{{\rm B\kern-.05em{\sc i\kern-.025em b}\kern-.08em
    T\kern-.1667em\lower.7ex\hbox{E}\kern-.125emX}}

\usepackage{makecell}
\usepackage{threeparttable}
\makeatletter  
\newif\if@restonecol  
\renewcommand\footnoterule{%
	\kern-3\p@
	\hrule\@width\columnwidth
	\kern2.6\p@}

\begin{document}

\newcommand{\tool}{{SceneDroid}\xspace}
\newcommand{\graph}{{SceneTG}\xspace}
\newcommand{\ling}[1]{\textcolor{red}{00: #1}}
\newcommand{\zxy}[1]{\textcolor{orange}{#1}}
\newcommand{\sen}[1]{\textcolor{blue}{#1}}

\title{Scene-Driven Exploration and GUI Modeling for Android Apps
}


\author{
\IEEEauthorblockN{Xiangyu Zhang\IEEEauthorrefmark{1},
Lingling Fan\IEEEauthorrefmark{1}\IEEEauthorrefmark{4},
Sen Chen\IEEEauthorrefmark{2},
Yucheng Su\IEEEauthorrefmark{3},
Boyuan Li\IEEEauthorrefmark{1}
}
\IEEEauthorblockA{\IEEEauthorrefmark{1}DISSec, NDST, College of Cyber Science, Nankai University, China}
\thanks{\IEEEauthorrefmark{4} Lingling Fan is the corresponding author (linglingfan@nankai.edu.cn).}
\IEEEauthorblockA{\IEEEauthorrefmark{2}College of Intelligence and Computing, Tianjin University, China}
\IEEEauthorblockA{\IEEEauthorrefmark{3}Chaitin Technology, Alibaba Group, China}
}

\maketitle

\begin{abstract}
Due to the competitive environment, mobile apps are usually produced under pressure with lots of complicated functionality and UI pages. Therefore, it is challenging for various roles to design, understand, test, and maintain these apps. The extracted transition graphs for apps such as ATG, WTG, and STG have a low transition coverage and coarse-grained granularity, which limits the existing methods of graphical user interface (GUI) modeling by UI exploration. To solve these problems, in this paper, we propose \tool, a scene-driven exploration approach to extracting the GUI scenes dynamically by integrating a series of novel techniques including smart exploration, state fuzzing, and indirect launching strategies. We present the GUI scenes as a scene transition graph (SceneTG) to model the GUI of apps with high transition coverage and fine-grained granularity. Compared with the existing GUI modeling tools, \tool has improved by {168.74\%} in the coverage of transition pairs and {162.42\%} in scene extraction. Apart from the effectiveness evaluation of \tool, we also illustrate the future potential of \tool as a fundamental capability to support app development, reverse engineering, and GUI regression testing.
\end{abstract}

\begin{IEEEkeywords}
Android app, Scene-driven exploration, GUI exploration, GUI modeling
\end{IEEEkeywords}

\section{Introduction}
Mobile applications (apps) are indispensable for daily life~\cite{chen2019storydroid}. Excessive demand also means that people have higher requirements for these apps, therefore, they are usually developed under pressure with more complex functionalities and UI pages. Every coin has two sides. It is challenging to design, understand, test, and maintain these apps for different roles such as product manager, designer, developer, and maintainer. To mitigate such a problem and to help understand these complex apps, app abstract and graphical user interface (GUI) modeling have been used to realize apps by levering UI exploration~\cite{azim2013targeted,yang2018static,chen2019storydroid,lai2019goal,chen2022automatically,zhang2023web}. 
Many different approaches to GUI modeling are raised gradually such as activity transition graph (ATG)~\cite{azim2013targeted, fan2018efficiently}, window transition graph (WTG)~\cite{yang2018static}, and screen transition graph (STG)~\cite{lai2019goal}.

Although static and dynamic methods are available for UI exploration, there are two significant issues that have not been dealt with yet. \textit{(1)} it is challenging to construct a relatively complete $*$TG.\footnote{We use $*$TG to present these existing transition graphs.} Due to numerous implementations and various code styles, the static UI exploration is missing several transitions~\cite{yan2022comprehensive,chen2022automatically}. Besides, as some activities are too complex to fully explore or required complex inputs that cannot be completed automatically, the coverage may still be far from acceptable \cite{su2021benchmarking,choudhary2015automated,zeng2016automated}. \textit{(2)} The UI pages are more significant than the $*$TG structure since Android apps are event-driven with rich UI pages. The UI page is more helpful and intuitive for users to understand the app.

Under the situation, Chen et al.~\cite{chen2019storydroid} inspired by the conception of storyboard in the movie industry, proposed StoryDroid and automatically extracted storyboards for Android apps, which contains both ATG and rendered UI pages along with many other useful features such as UI components, the corresponding layout and logic code, method hierarchy. Another work StoryDistiller~\cite{chen2022automatically} is an extension of it \cite{chen2019storydroid}, which enhanced StoryDroid on both the ATG construction and UI page rendering by adding dynamic UI exploration. In other words, StoryDistiller is a hybrid solution to extract storyboards for apps with rich features for app abstract and GUI modeling with rich visible UI features.

\begin{figure}
\centering
\includegraphics[width=\columnwidth]{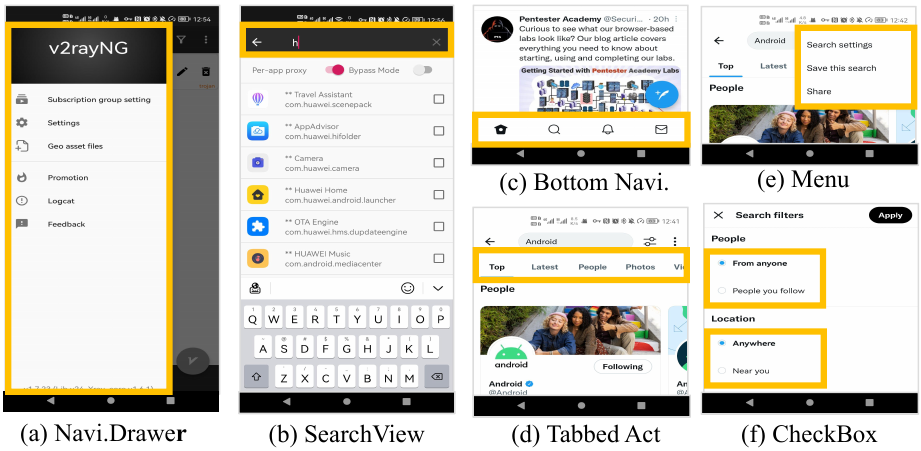}
\caption{Scene examples triggered by different UI components.}
\label{fig:Intro}
\vspace{-0.3cm}
\end{figure}

However, StoryDistiller~\cite{chen2022automatically} still has shortcomings that obstruct understanding and realizing apps: \textit{(1)} The strategy of dynamic exploration is only to trigger each interactive UI component on the rendered activity, missing many deep-level interactive UI components. The simple strategy inevitably lost a lot of transition pairs. \textit{(2)} The extracted $*$TG is coarse-grained. In addition to the $*$TG, many other GUI ``scenes'' can be triggered in activity as shown in~\Cref{fig:Intro}, leading to the creation of numerous new UI pages containing new functionalities. An urgent need for a fine-grained GUI modeling solution exists. In fact, addressing the above-mentioned problems poses the following challenges: \textbf{C1:} \textbf{Reasonable UI Granularity.} Achieving a reasonable UI granularity is challenging when seeking to define app UI updates, as we must preserve key UI information while avoiding the recording of excessive unnecessary states. An overly coarse granularity may lead to misjudgments of UI states, adversely affecting test results, while an excessively fine granularity may generate a multitude of redundant states, hindering testing efficiency. Consequently, identifying an appropriate granularity balance to achieve efficient and accurate UI update recognition is a key challenge. \textbf{C2:} \textbf{Launching Activity.} During the dynamic exploration of Android apps, enhancing the ability to launch activities is a key challenge. Android apps typically comprise multiple activities, which are the core components of the app, responsible for displaying various user interfaces and handling user interactions. However, during the dynamic testing process, some activities may not be easily triggered, as they might require specific user input or a particular application state. Furthermore, certain activities might only be triggered under specific conditions, rendering the dynamic exploration process potentially unable to cover all possible activities.

To this end, in this paper, we propose \tool, a scene-driven exploration and GUI modeling approach, which leverages a smart exploration to dynamically extract the GUI scenes.\footnote{In this paper, a \textbf{scene} is defined as the UI page that is triggered by interactive UI components of the activity $A$, whose layout is different from that of $A$. Such new scenes may be rendered as the current activity $A$ with new views, a new fragment of $A$, or a new activity.} Specifically, to address C1, \tool proposes a scene recognition method that considers the hierarchical structure of components on the UI page and ignores minor changes that may lead to layout changes, thus identifying unique scenes. \tool constructs a finer GUI model based on scenes, called the Scene Transition Graph (SceneTG). To address C2, \tool designs an exhaustive exploration strategy to explore all scenes of an app and interact with as many interactive UI components as possible. \tool also introduces state fuzzing techniques to improve scene transition coverage. Most importantly, \tool designs an indirect launch strategy that leverages already explored activities to indirectly launch activities that Inter-Component Communication (ICC) messages failed to launch.

To demonstrate the effectiveness of \tool, we conducted comprehensive experiments. To evaluate the scene identification ability of \tool, we run it on 10 self-developed apps containing different types of interactive UI components that can trigger new scenes, results show that \tool can recognize all the preset scenes. 
We further compared \tool with 4 state-of-the-art GUI modeling tools to evaluate the effectiveness on 100 apps. The results demonstrate that the \tool surpasses other existing tools in terms of the number of transition pairs (30.25 on average) and scenes (22.93 on average). With improvements of 168.74\% in transition pair coverage and 162.42\% in scene extraction, \tool has significantly enhanced its performance. 
In addition, we also conducted an ablation study to evaluate the contribution of each strategy employed by \tool. The result indicates that the Indirect Launching strategy is the most contributing one, achieving an average improvement of 15.59\% in terms of activity exploration, 47.02\% improvement in scene exploration, and 35.08\% improvement in transition pair extraction.
As \tool serves as a fundamental tool for app exploration, we also discussed some applications based on \tool such as regression testing and UI-based testing.

In summary, we made the following contributions.
\begin{itemize}
\item We propose \tool, which is a novel approach leveraging a set of new techniques to construct the fine-grained app UI model by defining the scene transition graph (SceneTG). It can handle both open-source and closed-source apps.

\item \tool proposes a smart exploration algorithm, which mainly includes three strategies of exhaustive exploration, state fuzzing, and indirect launch method. These techniques improve the depth of exploration and the completeness of the SceneTG. 

\item Our comprehensive experiments demonstrate the effectiveness of \tool in app exploration and UI modeling compared with existing tools. Moreover, our experiments indicate the indirect launch strategy is the most contributing one to improving UI modeling.

\item This is a fundamental work providing a novel UI modeling method for apps, which facilitates future work in the reverse analysis of app structure, design and guidance of app development, creation of regression testing tools, etc. We have released \tool and the experimental dataset on \url{https://github.com/SceneDroid/SceneDroid}.

\end{itemize}

\begin{figure*}[!htb]
\centering
  \includegraphics[width=0.9\textwidth]{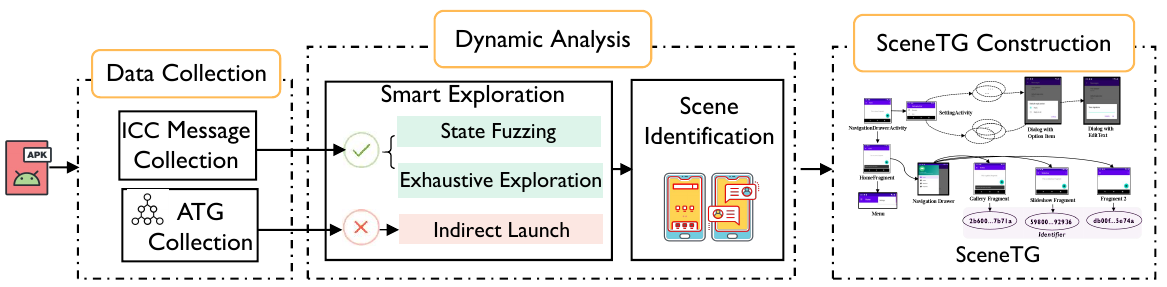}
  \caption{An overview of \tool.}
  \label{fig:overview}
  \vspace{-0.3cm}
\end{figure*}

\section{Background}
\subsection{Android Activity and Fragment}
        The Activity is the keystone of all Android apps. A component that contains a user interface primarily for user interaction. Android Fragment is a type of view that can be embedded in an activity. An activity can contain more than one Fragment, and a Fragment can also be reused in multiple activities, which can adapt to devices with different resolutions and make screen space utilization more reasonable. Like mini-activity, Fragment has its own layout and lifecycle \cite{li2017data}.

\subsection{Android UI Components}
        Android provides a large number of UI components\ \cite{xiao2019iconintent}\ that can be used flexibly to have a grandstand view of the app's functionality. For example, TextView is mainly used to display a text message on the current page. Button is an essential UI component used to interact with the users. Button objects can receive user-clickable events. ImageView and ImageButton are UI components available for displaying icons. In addition to these common and basic types, other types of UI components are usually used to enrich the user interface. For example, Menus are used in most apps to deliver user actions and some options. The menus are often laid out with important options that allow changes to be made to the environment variables and environment data that the apps depend on. The navigation drawer is one of the most general effects in Material Design which can hide some menu options on the left of the top app bar. It can display the main navigation items of the app. AlertDialog and ProgressDialog can pop up dialogs on the current page.

\section{Approach}
Fig.~\ref{fig:overview} shows the overview of \tool, which consists of three main parts: data collection, dynamic analysis, and scene transition graph (SceneTG) construction. \tool takes an APK file as input and outputs a visual SceneTG and other related parsing results such as the corresponding screenshot for each scene and its corresponding layout files. 
The data collection module collects the Inter-Component Communication (ICC) message for activity launching to facilitate dynamic analysis and the activity transition graph (ATG). The dynamic analysis module runs the apps by employing the Smart Exploration algorithm and identifies new scenes. The SceneTG Construction module takes the outputs of dynamic analysis to generate the SceneTG, including the screenshot of each scene and the scene transitions.

\subsection{Data Collection}
    
    The goal of data collection is to provide the dynamic exploration module with as much information as possible, 
    {including the ICC messages for direct activity launching and ATG for indirect activity launching,}
    so as to improve the efficiency and effectiveness of dynamic analysis.

\subsubsection{\textbf{ICC Message Collection}} 
Android enables activity launching via console interfaces, with some requiring extra data. ICC messages, mainly Intent objects with data items, launch target activities. Generating ICC messages entails identifying Basic Attributes and Extra Parameters, found in intent-filters or Java code. Extra Parameters provide necessary specific data for successful launching. Comprising basic structures like String, Char, and Boolean, we generate data according to types to populate the Extra Parameter. The resulting Basic Attribute and Extra Parameter form ICC messages, used for activity launching and supplied to the dynamic analysis module.\par

\subsubsection{\textbf{ATG Collection}} 
Activity Transition Graph (ATG) is also one of the important features for app exploration, which states the transition relations between different activities. 
Lots of studies have been proposed to construct ATGs \cite{azim2013targeted,chen2019storydroid,yan2020multiple,yan2022iccbot,chen2022automatically}, and we use them to collect the initial ATGs for further analysis.  

In this paper, ATG is mainly used to guide \tool in the following dynamic analysis, especially when the activities fail to be launched directly with ICC messages, ATG can facilitate the exploration by providing the precursor activity for launching. Besides, ATG will be augmented by dynamic analysis and acts as the basis to construct the SceneTG.

\subsection{Dynamic Analysis}
 Based on the collected data, the dynamic analysis aims to exhaustively explore the scenes within the apps and identify new scenes and scene transitions during exploration. 
\subsubsection{\textbf{Smart Exploration}}
    Smart exploration focuses on obtaining as many different scenes as possible within an app. To achieve it, three strategies are designed: (1) \textit{{State fuzzing}}; (2) \textit{Exhaustive exploration of each activity}; and (3) \textit{Indirect launching for failed activities}, where different strategies are used in different stages.
    Specifically, given an app, \tool first tries to launch each activity based on the obtained ICC messages, the target activity is launched successfully, and the first two strategies are used to explore each activity exhaustively. If the activity fails to be launched, \tool will employ the third strategy to indirectly launch activities first and then continue using the first two strategies to explore activities. 
    Details are described as follows.
    
\noindent $\bullet$ \textbf{{State fuzzing}}.
    Since some activities contain UI components that users can interact with, however, would not trigger a transition to other scenes including EditText, CheckBox, Switch Button, etc. These kinds of components would not cause scene transition, however, may change the execution path of the app and thus potentially explore more states and scenes.
    Motivated by this, before operating on the interactive components that would trigger new scenes (e.g., Button, ImageButton, MenuButton), we proposed to employ the {state fuzzing} strategy first.
    
    Specifically, we consider employing fuzzing on 3 types of such non-transitive UI components: EditText, CheckBox, and Switch Button.
    For EditText, since some apps require user input to proceed to the next step, such as adding new items or searching the interface, we need to determine the format or some specific inputs that the component requires users to enter. 
    To achieve it, we first dump the Component Tree (i.e., UI layout) of the current activity, and extract the attributes of EditText, such as className, resource-id, and bounds. Since the dynamically obtained layout does not contain information about the required type of user input in terms of EditText, we use the extracted attributes to match the component declared in the source layout files, and obtain the required type of string (declared in inputType). We have summarized text, number, phone, date, time, and EmailAddress as common inputType. According to different input types, \tool will randomly generate a correctly formatted string and fill it into the specific EditText. 
    For CheckBox and Switch Button, we can directly identify them by the component type in the layout file. These two kinds of components have two states, checked or not checked (open or close, respectively). We can set them easily by clicking them.
    
    When there are multiple types of the aforementioned non-transitive UI components on a single activity, to explore potential new scenes, we go through all the possible combinations to form an initial state for the next strategy (i.e., exhaustive exploration). For example, if an activity contains all these 3 types, i.e., EditText has two values (``fill in'' or ``blank''), similarly, CheckBox has values of ``checked'' or ``not checked'', and Switch Button has values of ``open'' or ``close''. \tool will consider all the combinations of them and finally generate $2^3=8$ initial activity states for further exploration.

 \noindent $\bullet$ \textbf{Exhaustive exploration.}
    From a high level, \tool employs a breadth-first strategy at the Activity level, while exploring scenes on a specific activity, \tool uses a depth-first strategy, aiming to explore as many scenes within the activity.
    Therefore, based on each generated initial activity, \tool extracts all the actionable components according to the attribute ``clickable=true'' of each component in the dumped layout file, such as Button, ImageButton, CheckBox, ImageView, and RadioGroup. 
    It combines these actionable components into an exploration queue and takes one component at a time from the queue to interact with.
    When a new scene associated with the current activity is identified, \tool will record its layout file, screenshots, and experienced components.
    Besides, \tool iteratively performs this exploration process on the scene and records the scene transition relation as $scene_1\overset{e, c}{\rightarrow}scene_2$ where $e$ and $c$ represent the event and component triggering this transition, respectively.   
    If it does not reach the new scene or reaches a visited scene, it returns to the previous scene and interacts with the next component.
    In addition, during exploration, the current activity $A$ may transit to a new activity $B$ by operating on specific components (i.e., activity transition), \tool will rollback to $A$ and continue exploring other scenes within $A$. Such activity transitions (i.e., $A\overset{e, c}{\rightarrow}B$) are also recorded to augment the static ATG and are further used to help exploration and SceneTG construction.

\begin{figure}
\centering
\includegraphics[width=0.99\columnwidth]{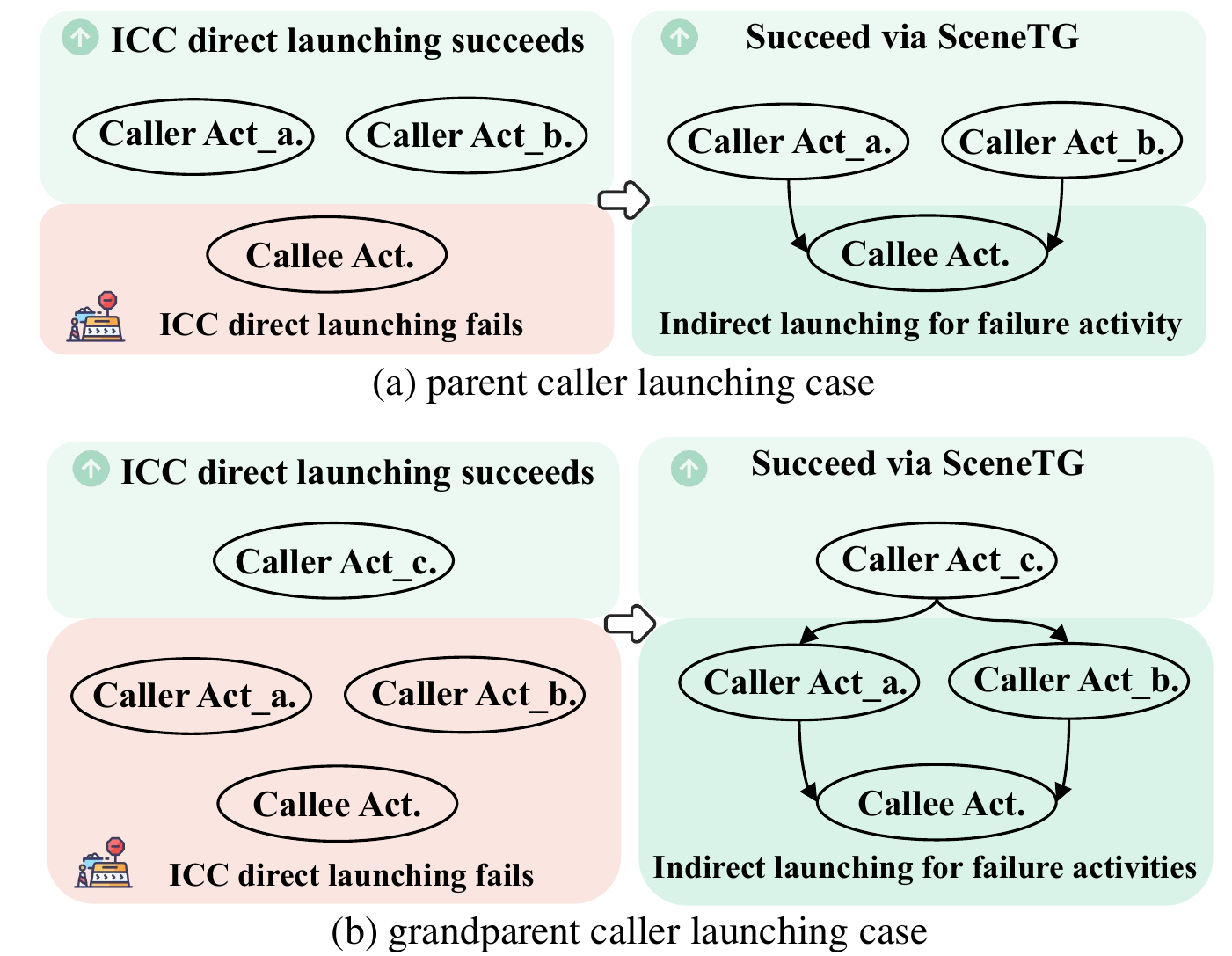}
\caption{Cases of indirect launching for failed activities.}
\label{Fig:IndirectFig}
\vspace{-0.3cm}
\end{figure}
    
    \noindent  $\bullet$ \textbf{Indirect launching for failure activities.}
    Due to the inconsistency of activity declaration between the app implementation and the AndroidManifest.xml file or incorrect static ICC messages, some activities may not be launched successfully with ICC messages.
    \tool will find the upstream caller activity as a bridge to indirectly launch the target activity, by utilizing the SceneTG that has been constructed so far. 
    For example, in Fig. \ref{Fig:IndirectFig}(a), when \textit{Callee Act.} failed to be launched with ICC messages, \tool will find the caller of it from ATG, i.e., \textit{Caller Act\_a} and \textit{Caller Act\_b}, both of which can be used to indirect launch \textit{Callee Act.}
    Note that ATG is dynamically augmented and updated during exploration, here, we use the latest ATG to ensure the successful launch of the target activity.
    Specifically, if an activity $act_{des}$ failed to be directly launched with ICC messages, \tool will traverse the ATG and find the caller activity of $act_{des}$, i.e., $act_{src}$, where $act_{src}\rightarrow act_{des}$. After that, we will try to launch $act_{src}$ with ICC messages, if it is successfully launched, we then use the event (i.e., action) that triggers such an activity transition and operate on it to launch $act_{des}$.
    To extract the events triggering the specific transition, we use the maintained ATG which contains the transition relation between different activities together with the events and components that trigger such relation, i.e., $act_{A}\overset{e, c}{\rightarrow} act_{B}$.

    However, there may be cases that the direct caller activity $act_{src}$ cannot be launched, either. Therefore, we obtain a list of caller activities as the candidates to launch $act_{des}$.
    For example, in Fig. \ref{Fig:IndirectFig}(a), the direct callers of the failed activity (i.e., \textit{Caller Act\_a} and \textit{Caller Act\_b}) both failed to be launched, we thus iteratively find the caller of the failed ones and finally launched \textit{Callee Act.} via launching \textit{Caller Act\_c}.
    Once the target activity ($act_{des}$) is directly launched by one of the caller activities, we stop this process and employ the two strategies above (i.e., {state fuzzing} and exhaustive exploration) to explore this activity and the associated scenes.
    If all the candidate caller activities fail to launch $act_{des}$ indirectly, we temporarily move it to the end of the exploration queue and continue exploring other activities. For $act_{des}$, we update ATG and launch it iteratively by traversing it.
    
    \begin{algorithm2e}[t]
	\setcounter{AlgoLine}{0}
	\caption{Smart Dynamic Analysis}
	\label{algo:explore}
	\DontPrintSemicolon
	\SetCommentSty{mycommfont}
	\KwIn{$act_{all}$: All activities with ICC messages in the app; $ATG$: Activity transition graph.}
	\KwOut{$S$: All scenes explored within the app}
	$S \gets \varnothing$\;
	\ForEach{$act, icc$ $\in$ $act_{all}$}{
        \If{Success($act$, $icc$)}{
            ExploreAct($act$)
        }
        \Else{ \tcp*[h]{Failed to launch $act$.} \;
            $act_{caller}$ = IndirectLaunch($act$, $ATG$)\;
            \If{$act_{caller} \neq Null$}{ 
                ExploreAct($act$)\;
            }
            \Else{
                \tcp*[h]{No such a caller act that can launch $act$, then $act$ is added to the queue for a second launch} \;
                $act_{all} \gets act_{all} \cup act$\;
            }
        }
        
    }
        \SetKwFunction{ExploreAct} {ExploreAct}
        \SetKwProg{Fn}{Function}{:}{}
        \Fn{\ExploreAct{$act$}}{
           $States \gets$ Fuzzing($act$)\;
           \ForEach{$st \in States$}{
                $S$ $\gets$ ExhaustiveExplore($st$)\;
           }
        }
        \Return{$S$}\;
    \end{algorithm2e} 
    
    Algorithm \ref{algo:explore} depicts the whole process of smart dynamic exploration, which employs the three strategies alternatively. 
    The input is all the activities with ICC messages for launching ($act_{all}$), and \tool outputs the scenes ($S$) explored by using the three strategies.
    Specifically, $S$ is first initialized as empty and will be gradually augmented during exploration. For each activity $act$, we first try to directly launch $act$ by using the associated ICC message. If $act$ is launched successfully, we continue to employ the fuzzing strategy and exhaustive exploration on it by calling the method \texttt{ExploreAct} (Lines 3-4). 
    In the activity exploration process (Lines 11-14), we first employ the fuzzing strategy to generate different initial states ($States$) for $act$ (Line 12), and for each state, we start exhaustive exploration (Lines 13-14) and store the explored scenes in $S$.
    However, if $act$ fails to be launched, we employ the indirect launch strategy to identify the caller activity of $act$ that can indirectly launch it based on the latest ATG (Line 6). If there exists such a caller activity $act_{caller}$, we utilize it to transit to $act$, and continue to employ the fuzzing strategy and exhaustive exploration on it (Lines 7-8). Otherwise, $act$ is added to the exploration queue for a second launch (Lines 9-10), because the ATG is dynamically updated during exploration, the augmented ATG later may be able to launch $act$. Therefore, we employ it to maximize the possibility of launching each activity. If the ATG is not augmented after an exploration round, we stop re-launching the failed activities, and stop the whole process and return $S$ (Line 15).

\subsubsection{\textbf{Scene identification}}

Since the goal of \tool is to construct a relatively complete UI model consisting of different types of fine-grained UI states, i.e., \textbf{scene}, we proposed a scene identification method, aiming to identify the unique scenes by abstracting and modeling the UI pages in a fine and suitable manner, so as to avoid keeping exploring duplicated scenes. 
The scenes identified by \tool include activity, fragment, drawer (e.g., Top/Bottom/Side navigation drawer), dialog, menu, checkbox, spinner, picker, floating action button, etc., some are shown in Fig.~\ref{fig:Intro}.

\begin{figure}
    \centering
    \includegraphics[width=0.75\columnwidth]{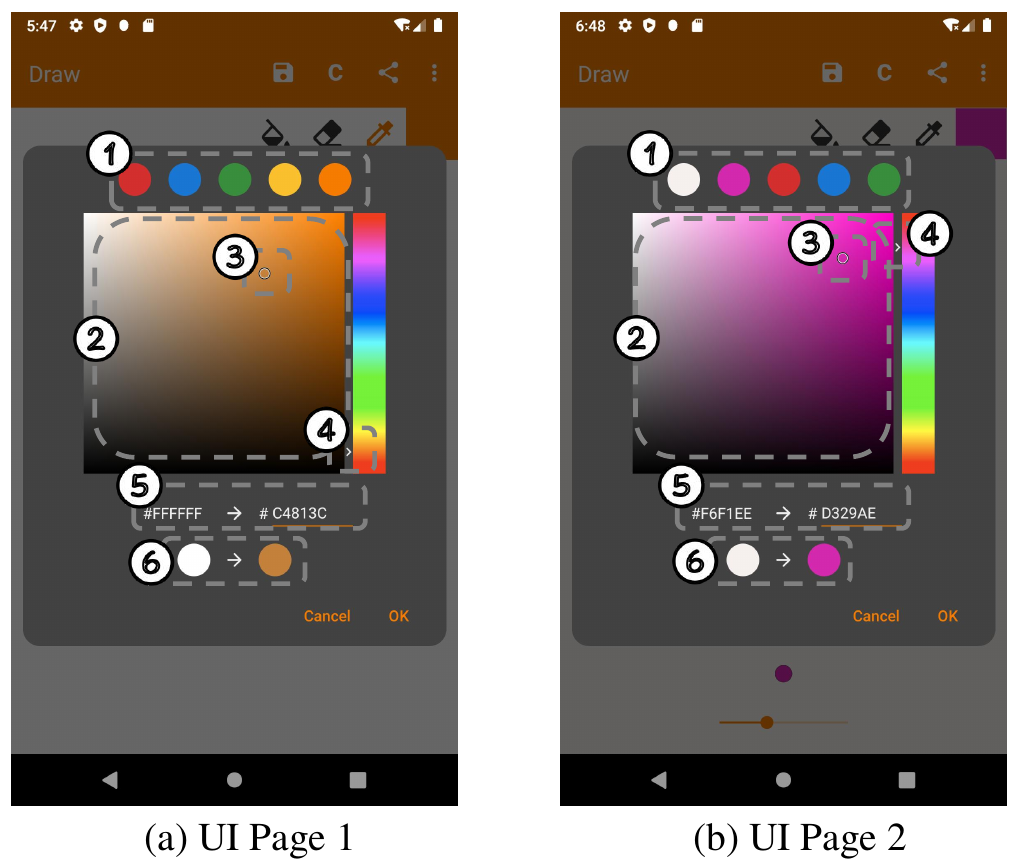}
    \caption{Two UI pages in the app \texttt{Simple Draw Pro}.}
    \label{Fig:palette}
    \vspace{-0.3cm}
\end{figure}

Specifically, for each explored UI page, we aim to generate a unique identifier based on the layout dumped dynamically as an abstraction of the UI page. If the identifiers of two UI pages are the same, we regard them as the same scene, otherwise, two scenes are both recorded.
To avoid maintaining a massive number of scenes with subtle changes, and model the UI page in a fine and proper grained, we consider abstracting a UI page based on the hierarchy of components on it, the unique ID of each component (i.e., \textit{resource-id} in the layout file), the type of the components (i.e., \textit{class}), and the package it belongs to (\textit{package}).
These attributes preserve the number and the type of components, as well as their hierarchy, meanwhile omitting the subtle changes (such as the text change and color change) which would not cause layout changes but may lead the exploration to a dead end.
For example, in Fig. \ref{Fig:palette}, this is a simple drawing app that produces several UI changes when the user selects different brush colors.
Since no matter how the values of these UIs change, it is just about the color selection with different values and would not cause an impact on the structure, we thus consider them as the same scene.

In detail, for each UI page, we first dump the layout file which contains all the components and their attributes (e.g., resource-id, text, class, package, clickable), and each node represents a component.
We then record the hierarchy of all the components and start a Breadth-First traversal to obtain the component sequence as a list. Note that, since \tool dumps the layout structure directly from the UI page, which may introduce the UI of other packages, such as the UI of the status bar or the UI of the input method when it pops up. The UI with these non-target packages will interfere with the judgment of the current UI page, but directly ignoring them may lead to missing new scenes. Therefore, we decided to discard the non-target package UI in \tool,
and only considered the nodes that belong to this app by matching the package names. For each component in this sequence, we extract the value of three attributes as the unique identifier of it, i.e., \textit{resource-id}, \textit{class}, \textit{package}. We then concatenate these three attribute values and use the MD5 hash algorithm \cite{rivest1992md5} to generate a hash value for the component. If the type of the current node is an adapter view, we will use the information of the view it is really bound to generate the identifier for it.
After obtaining the hashed values for all the components, we concatenate them in sequence and use the same hash algorithm to generate a unique identifier for the UI page.

Note that, since the detailed contents in adapter views (e.g., ListView or RecyclerView) at runtime are unknown and these adapter views are essentially just repetitive views being populated according to the ListApdapter\cite{Widget} . 
While \tool focuses on the structure of the views obtained from the ListApdapter, it only needs to fetch the first view in the adapter view to learn the structure of the other ones. Only the first child view of adapter views counts for scene identification.

\begin{figure}
    \centering
    \includegraphics[width=0.99\columnwidth]{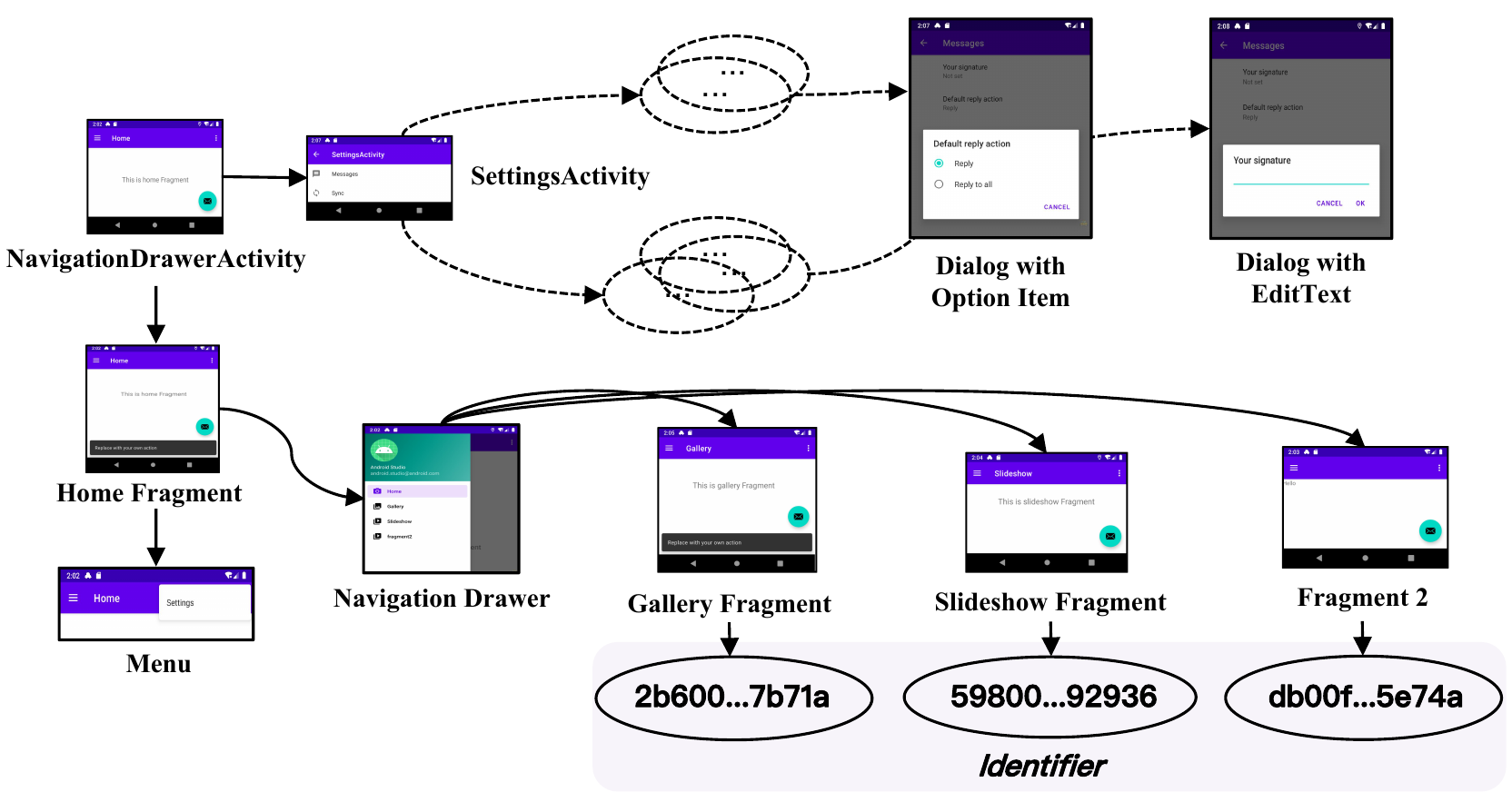} 
    \caption{Example of a SceneTG constructed by \tool.}
    \label{fig:stg}
\end{figure}

\subsection{SceneTG Construction}
To reflect the overall UI states of an app in the runtime, we construct the  based on the identified scenes and their transitions during dynamic exploration (as shown in Fig. \ref{fig:stg}). We highlight that apart from the scene transitions, \tool also can provide the corresponding real UI page for each identified scene. The SceneTG attached with real UI pages indeed aids users in understanding the apps.
SceneTG's fine-grained UI model can be used to contribute to improving the performance of existing work including UI testing, regression testing, competitive product analysis, etc.

\section{Effectiveness Evaluation}

To evaluate the effectiveness of \tool, we aim to conduct the experiments by answering the following research questions.
\begin{itemize}[leftmargin=*]
    \item \textbf{RQ1}: Can \tool accurately recognize new scenes that contain different types of new UI views?
    \item \textbf{RQ2}: Can \tool outperform existing UI exploration tools in terms of transition relation extraction and scene exploration?
    {\item \textbf{RQ3}: How much do the different strategies of \tool contribute to enhance UI exploration?}
\end{itemize}

\label{sec:dataprepare}

\subsection{RQ1: Scene identification}
\subsubsection{\textbf{Setup}}

To investigate whether \tool can effectively identify different types of scenes in the apps, we self-developed 10 apps as our ground-truth benchmark, covering different types of views for UI pages including Drawer, Menu, Dialog, Spinner, Picker, etc.
In order to make the benchmark apps more representative of real-world apps, we also add more features and complexity to them with different numbers of activities. Since Android Studio provides numerous code templates that follow the best practice of Android app design and development, to develop apps that are compliant with the latest Material Design principles and reflect the latest Android app features, we utilize the templates provided by Android Studio to create new application modules, various activities, or other specific Android project components. Some templates provide initial code for typical environments, such as drawer navigation bars or login pages, which reflect the latest Android app features. As shown in Table 1, the 10 apps we develop consist of many features, varying the number of activities with multiple types of scenes. Moreover, they are implemented with different transitions from Activity to Activity, Activity to Fragment, Fragment to Activity, and Fragment to Fragment, as the rich transition logic that is inserted into the apps.

Based on the dataset above, we conducted the experiment to evaluate the effectiveness of \tool in scene identification. To validate the accuracy of \tool, we need to determine the number of activities, scenarios, and transition relations for each program. We use the number of activities declared in the AndroidManifest.xml file as the basis and manually validate the number of scenes and transition pairs identified by \tool for each app. We set a timeout of 15 minutes for the analysis phase and 30 minutes for the dynamic analysis for each app in the dataset.

\begin{table}
\caption{Ten self-developed benchmark apps with different features, activities, transition pairs, and scenes.}
\begin{center}
\begin{tabular}{|c|p{3cm}|c|c|c|}
\hline
\textbf{ID} & \textbf{Feature} & \textbf{\#All\_Acts} & \textbf{\#Pairs} & \textbf{\#Scenes} \\
\hline
1 & Basic Act + Fragment + Dialog + Switch Button & 8 & 23 & 17 \\
\hline
2 & Basic Act + Menu & 8 & 18 & 15 \\
\hline
3 & Navi. Drawer Act + Fragment & 9 & 24 & 22 \\
\hline
4 & Navi. Drawer Act + Fragment + Menu & 8 & 21 & 19 \\
\hline
5 & Bottom Navi. Act & 8 & 13 & 13 \\
\hline
6 & Bottom Navi. Act + Menu & 3 & 19 & 19 \\
\hline
7 & Bottom Navi. Act + Fragment + EditText & 3 & 15 & 14 \\
\hline
8 & Tabbed Act + Menu + Spinner + Picker. & 6 & 14 & 11 \\
\hline
9 & Tabbed Act + Bottom Navi. Act + Menu + Floating Action Button & 3 & 16 & 11 \\
\hline
10 & Navi. Drawer Act + Fragment & 1 & 6 & 9 \\
\hline
\end{tabular}
\label{tab:benchmark_apps}
\end{center}
\vspace{-0.3cm}
\end{table}

\subsubsection{\textbf{Result}}
    The result indicates that \tool can extract all the activities, scenes, and transition pairs in the 10 benchmark apps, shown in Table \ref{tab:benchmark_apps}. 
    \tool performed well not only on simple apps composed of activities and fragments but also on complex apps, as displayed in app 4 and app 9. These complex combinations of features are frequently used in industrial environments. 
    In the following RQ2, we will show in detail the strengths and weaknesses of \tool compared to others, especially in apps with complex components. 
    The reason for achieving such excellent results is that \tool leveraged a combination of three smart strategies. These strategies are not used in isolation or stacked repeatedly; rather, the organic combination achieves good results. 
    In the following RQ3, we will conduct an ablation study to comprehensively evaluate the impact of each strategy on the tool's exploration capability.
    {The SceneTG constructed by \tool can indeed build a more fine-grained UI model. We also manually verified the reachability of all the paths explored by \tool, and all of them are feasible in the 10 benchmark apps. 
    }

\smallskip
\noindent \fbox{
	\parbox{0.95\linewidth}{
	\textbf{Answer to RQ1:} The experimental results show that \tool can extract all activities, scenarios, and transition pairs in the 10 ground-truth benchmark apps. \tool can accurately recognize new scenes that contain different types of new UI views. 
	}
}

\subsection{RQ2: Scene exploration}
\subsubsection{\textbf{Setup}}
To evaluate the capability of \tool in scene exploration, we randomly downloaded 50 closed-source apps from Google Play Store\cite{googleplay} and 50 open-source apps from F-Droid\cite{fdroid} as the evaluation subject to investigate the effectiveness of \tool in real-world apps.
    Based on the dataset, we compared \tool with four state-of-the-art UI modeling tools: GoalExplorer \cite{lai2019goal}, Gator \cite{rountev2017gator}, StoryDistiller \cite{chen2022automatically}, and ICCBot \cite{yan2022iccbot}. 
    We chose them as the baseline tools because they either have similar goals (StoryDistiller) to \tool or have similar transition results (GoalExplorer, Gator, ICCBot).
   Specifically, StoryDistiller utilizes a combination of dynamic and static methods to build the UI model of the app, which is with a similar goal to ours but with coarse-grained modeling.
    The other three tools are state-of-the-art tools that generate transition graphs. 
    GoalExplorer proposes a static parsing approach to build the Screen Transition Graph (STG). Note that, in the experiment, we used the latest released version of GoalExplorer\cite{ReSeSS} since the initial open-source version on Github is unavailable to compile and use due to missing essential dependencies. 
    Gator is also a mature static analysis suite for Android apps that can be used to build the Window Transition Graph (WTG).
    ICCBot is demonstrated as the state-of-the-art ICC resolution tool \cite{yan2022comprehensive}.  

    We separately run these tools on the 100 apps and set a timeout of 15 minutes for each app in the static analysis phase, because, for some closed-source applications, some static analysis tools can be time-consuming due to internal errors. 
    For the evaluation metrics, we use the number of explored activities, the number of explored scenes, and the number of UI transition pairs to evaluate the performance of each tool. 
    Since ICCBot generates ICC relation of the four major components of Android (i.e., Activity, Service, Content Provider, and Broadcast Receiver), while \tool focuses on the UI model construction. To make a fair comparison, we thus only consider the components related to UIs from the result file, i.e., activity and fragment. As for the number of transitions of ICCBot, we focus on four types of transitions: Activity to Activity, Activity to Fragment, Fragment to Activity, and Fragment to Fragment.

\subsubsection{\textbf{Result}}

\begin{figure}
    \centering
    \includegraphics[width=0.48\textwidth]{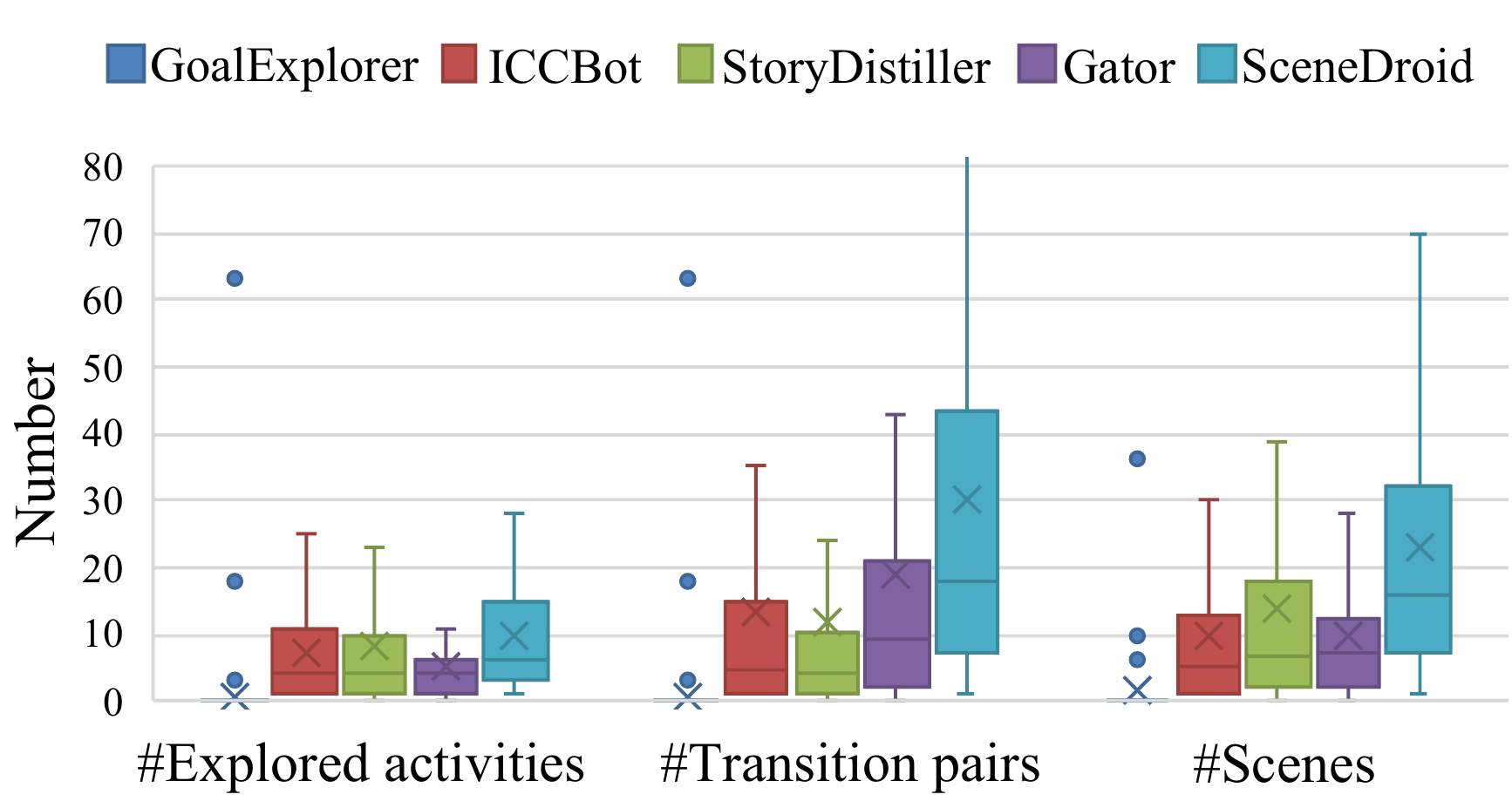}
    \caption{Comparison of \#Explored activities, \#Transition pairs, \#Scenes.}
    \label{Fig:RQ2comparison}
\end{figure}

    The comparison result of these tools is shown in Fig. \ref{Fig:RQ2comparison}. We can see \tool outperforms the other four tools in all metrics. On average, \tool extracts 30.25 transition pairs (0.81 in GoalExplorer, 13.52 in ICCBot, 12.03 in StoryDistiller, 18.68 in Gator, respectively), and in terms of the identified scenes, \tool achieves 22.93, which is twice of most other tools (1.63 in GoalExplorer, 9.53 in ICCBot, 13.83 in StoryDistiller, 9.95 in Gator, respectively).
 
    The reason for \tool's superior results is that 

    \tool introduces smart exploration, which is used to obtain the scenes during dynamic exploration, thus enabling the launch of activities even without using ICC messages. 
    It alleviates the limitations of existing tools that rely on the accuracy of ICC message extraction, effectively enhancing the activity coverage of \tool during the dynamic process. Smart Exploration also introduces the indirect launching phase for failure activities, which helps \tool to explore as many different scenes on an activity as possible. Moreover, fuzzing for EditText, CheckBox, Switch Button, etc., is an exclusive feature that enables \tool to interact with more components than other tools. 

    While StoryDistiller also adopted the idea of combining static and dynamic exploration to build UI models with UI screenshots, it does not perform well because \textit{(1)} StoryDistiller works with activity as a granularity. While it also tries to trigger each interactive component presented in the activity, it will only go to explore the ones that start the initial activity. Besides, it ignores the possibility of triggering components that will access a scene such as Fragment or Menu, where the newly emerging interactive components may trigger new scenes and new transition relations. \textit{(2)} StoryDistiller relies on ICC messages to launch the activity and cannot be assisted through the transition pairs obtained by the dynamic exploration process; \textit{(3)} StoryDistiller does not use fuzzing to increase interactions.

\begin{figure}
    \centering
    \includegraphics[width=0.48\textwidth]{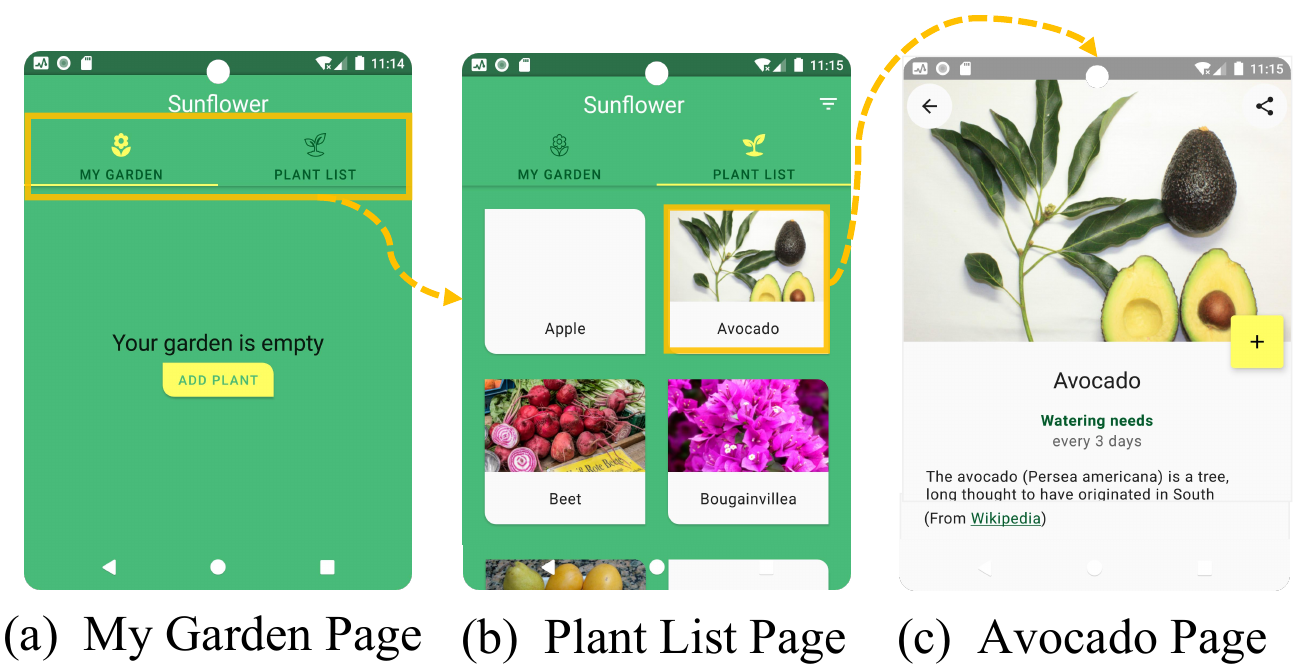}
    \caption{Example transition between Tabbed navigation UI.}
    \label{Fig:tabnavi}
\end{figure}

    {As the static analysis methods ignore many of the transition relations brought about by the presence of special components in the new view during analysis. Some components that can trigger the new scene exist in some new views (e.g., Navigation, Snackbar, and BubbleMetaData), while these static methods do not resolve the views, preventing them from triggering the new scene.}
    For example, none of the existing tools properly handle the transition pairs initiated by the Navigation components or navigated using Tabbed Navigation UIs, as shown in Fig. \ref{Fig:tabnavi}.
    Another example is in Fig. \ref{fig:stg}, they fail to properly analyze the transition pairs from the Navigation Drawer to the GallerFragment, SlideshowFragment, and Fragment2. Navigation is the interaction that allows users to navigate across, into, or back out from different content blocks in an app \cite{Widget}, which is introduced in Android 3.3.

    For StoryDistiller, it is based on the grain of activity, and discovering scenes containing Navigation components is beyond the capability of StoryDistiller. For ICCBot (which claims to be able to model Fragments) and GoalExplorer (which is optimized explicitly for Drawer) also fail to correctly discover the transition pairs generated by the Navigation component. This is because the API modeling of these tools failed to keep pace with Android evolution, and neither of them correctly modeled Navigation's API introduced in Android 3.3. 
    Specifically, in the Fragment-Aware Transition and Extraction phases, both of ICCBot and GoalExplorer only captured the APIs commonly used by FragmentManager. For example, when identifying the addition of a fragment, the APIs such as \texttt{add(Fragment, String)} are captured, while GoalExplorer only models the \texttt{DrawerLayout.openDrawer} API when dealing with the component Drawer. However, the APIs used for jumping between fragments in the Navigation component are 
    \texttt{Navigation.navigate(actionID)}
    and \texttt{Navigation.navigateUp()}. Therefore, they both fail to handle scenes and transition pairs based on the Navigation component properly. 

\smallskip
\noindent \fbox{
\parbox{0.95\columnwidth}{
\textbf{Answer to RQ2:} 
\tool extracts 30.25 transition pairs and 22.93 scenes on average, which significantly outperforms the existing tools (i.e., 1.63 in GoalExplorer, 9.53 in ICCBot, 13.83 in StoryDistiller, and 9.95 in Gator) in terms of scene exploration on our collected 100 apps.}}


\subsection{RQ3: Ablation study on different strategies}

\subsubsection{\textbf{Setup}}

To evaluate the contribution of different strategies (i.e., State Fuzzing, Scene Identification, and Indirect Launch strategy) in \tool for improving UI exploration, in this RQ, we conducted an ablation study. Specifically, we tested with a modified \tool based on the dataset in RQ2, which can disable a particular strategy alone and we can separately evaluate the three strategies. 
We ran \tool with different strategies disabled for each of the 100 apps and set a 15-minute runtime limit for each app during the analysis phase, the same setup as that in RQ2. Given that it may get into a duplicate state when some strategies are disabled, leading to extra time consumption, we also set a time limit of 30 minutes during the dynamic run phase. We evaluate the effectiveness of each strategy based on the number of explored activities, scenes, and UI transition pairs.

\begin{figure}
    \centering
    \includegraphics[width=0.5\textwidth]{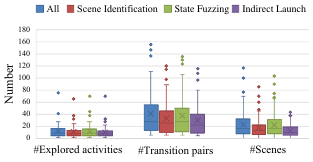}
    \vspace{-0.6cm}
    \caption{Contribution of each strategy.}
    \label{Fig:ablation}
\end{figure}

\subsubsection{\textbf{Result}}

The results of the ablation study are displayed in Fig. \ref{Fig:ablation}. The Indirect Launching strategy has the most impact on the test results of the tool, followed by the Scene Identification strategy. Specifically, in terms of activity exploration capability: the Indirect Launching strategy achieved an average improvement of 15.59\%  vs. 7.70\% in the Scene Identification strategy and 4.76\% in the State Fuzzing. Regarding the ability to explore Scenes, the Indirect Launching strategy provides an average 47.02\% improvement vs. 21.72\% in the Scene Identification strategy and 3,43\% in the State Fuzzing. The Indirect Launching strategy provided an average 35.08\% increase in the extraction of transition pairs. In comparison, the Scene Identification strategy provided an average of 19.86\% increase, and State Fuzzing provided an average of 7.89\% increase.

\begin{figure*}[t]
\centering
{
    \includegraphics[width=\textwidth]{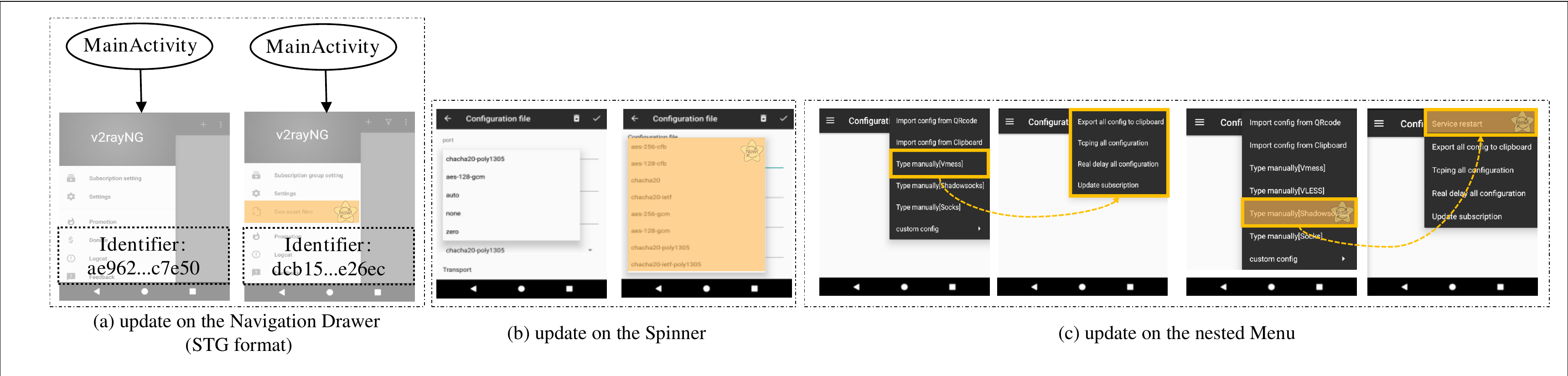}} 
    \caption{Fine-grained scene difference identification.}
\label{Fig:RQ3diff}
\end{figure*}

From the results, we can see that the \textbf{Indirect Launching strategy} contributes the most to the exploration capability of \tool. 
The possible reason is that, since Activity is the carrier for all scenes and transition pairs, once \tool is able to explore a new Activity that cannot be directly launched before, it would also explore a lot of new scenes and transition relations. 
It would be a practical strategy when the current static analysis techniques cannot fully construct the required context for launching activities correctly.

For the \textbf{Scene Identification strategy}, it brings a relative improvement to the scene exploration capability, proving that introducing the Scene Identification strategy is a justified choice. During the experiments, we found that disabling the Scene Identification strategy made it susceptible to repetitive scene exploration, which resulted in insufficient data. Once the tool gets stuck in repeated scene exploration, it is unable to exit automatically and thus fails to explore the whole application in a limited time. As in the case of the Scene Identification strategy described in the Approach section, apps like Simple Draw Pro get stuck because they cannot identify subtle scene differences. From the experimental results, with the Scene Identification strategy disabled, the tool was able to explore only seven different scenes. It was stuck in a repetitive exploration of the palette scenes. The introduction of the Scene Identification strategy proved to be feasible.

As for the \textbf{State Fuzzing strategy}, the boost is primarily because many apps contain UI components that users can interact with but do not directly cause scene transitions, including EditText, CheckBox, Switch Button, etc. However, these types of components can change the execution path of the application, making it possible to explore more new states and scenarios. In particular, many apps have scenes that require account passwords or search boxes, which may limit the exploration of scenes if not populated with appropriate data in the EditText component. Although the current State Fuzzing strategy improves the whole exploration, the improvement is relatively small, because some apps require legitimate input (e.g., specific account numbers and passwords) to be provided.

\smallskip
\noindent \fbox{
\parbox{0.95\linewidth}{
\textbf{Answer to RQ3:} 
{The Indirect Launching strategy has made the most significant contribution, with an average improvement of 15.59\%, 47.02\%, and 35.08\% in terms of activity exploration, scene exploration, and transition pairs extraction, respectively. The improvement is at least twice as effective as the Scene Identification strategy (7.7\%, 21.72\%, 19.86\%) and State Fuzzing strategy (4.76\%, 3.43\%, 7.89\%).}}}

\section{Future Applications and Discussion}
    \subsection{Future Applications}
    
    In this paper, we conduct fundamental work in UI exploration and fine-grained scene modeling, which can facilitate several follow-up research such as regression testing, and GUI testing for Android apps.

    \subsubsection{\textbf{Regression testing}}
    
    One of the meaningful areas of Android app testing is regression testing, as regression testing aids agile development in building quality apps. Moreover, related work shows that reusing test samples contribute to the efficiency of Android regression testing \cite{do2016redroid,sharma2019qadroid,peng2021cat}. Through experiments, we have demonstrated that \tool benefits from the high-precision UI model it builds and enables effective detection of modification scenes and components occurring in different app versions.
    By leveraging \tool, developers can focus more on testing the changed or added components or scenes, avoiding keeping testing on the previous functions.
    Besides, with the help of \tool, developers can write targeted test cases manually or using automated tools depending on the testing report. Goal-driven test case writing reduces the redundancy of testing and significantly saves the time required for testing. 

    We also conducted a pilot study to investigate whether \tool is capable of identifying fine-grained UI changes between different versions of the same app.
    Specifically, we randomly selected 30 apps in the dataset of RQ2 and collected the three latest minor versions\cite{Preston-Werner} of each app as the evaluation subject.
    As for UI changes (i.e., updates), we abstract the following two cases as updates: one is adding or deleting scenes, and the other is modifying components within the scene. We identify the UI changes by comparing the component tree of the two scenes (with the same execution path) of the two versions, \tool checks layer by layer whether any nodes have been added or deleted or the properties of the old nodes have been changed.
    Based on the dataset and the update localization method, we aim to investigate the number of scenes and transitions updated in the newer versions that are identified by \tool.

    As a result, \tool found 135 updates of scenes and 284 updates of transition pairs in 60 adjacent version iterations of 30 apps. On average, each version update introduces 1.50 scene variations and 3.20 variations of transition pairs, indicating that scene updates are relatively frequent during app evolution. 
    {Take the app V2Ray \cite{V2RayNG} as an example, which is a Material-Design-compliant web proxy application. We first discover an update of the $NavigationView$ on the $DrawerLayout$. As shown in Fig. \ref{Fig:RQ3diff}, from version 1.4.0 to 1.5.0 of the app, V2Ray was updated to support the custom functional modules (i.e., Geoip and Geosite). This feature update visually reflects the difference in scene, with a new entry for ``Geo asset files'' in $NavigationView$. The identifier of this scene is also changed from ``ae96...7e50'' to ``dcb1...26ec``,
    which can also be found visually in the SceneTG (Fig. \ref{Fig:RQ3diff}(a)). \tool then applies the location algorithm mentioned above to find an additional node in the tree with a resource-id of ``com.v2ray.ang:id/user\_asset\_setting'', thus pinpointing the range affected by the update.}
    
    {In addition to the updated case of NavigationView on DrawerLayout, we also found an updated case on Spinner in V2Ray. As shown in Fig. \ref{Fig:RQ3diff}(b), the new version of V2Ray adds support for various forms of encryption, including ``chacha20'' and ``aes-256-gcm'', an update option that would be ignored if it were a traditional ATG or STG constructed by GoalExplorer. On the other hand, the SceneTG defined by \tool detects this granularity update very well.}
    {\tool can also find the updated scene on Menu. This Menu update case is unusual because it happens on a nested Menu (as shown in Fig. \ref{Fig:RQ3diff}(c)). Version 1.3 added support for the VLESS protocol compared to version 1.2, so there is a new entry point on the Menu imported by the protocol. \tool observes the scene update on this first Menu; however, it can be seen that there is also a custom configuration option. Clicking on this custom configuration option, \tool finds a second Menu, adding in version 1.3 the ability to restart all services, which needs to be triggered in the second nested Menu. This UI update could not be found if only the general activity or activity to Menu level granularity was created. }
    Due to the fine granularity of the scene and the exhaustive exploration strategy introduced by \tool, UI updates in the nested Menu can be discovered accurately.
    \tool can identify the fine-grained UI changes based on graphs of SceneTG between multiple versions of the same app.

    \subsubsection{\textbf{Android UI testing}}
    Prior research has shown that even with the current state-of-the-art Android GUI testing kits, the activity coverage is still not high \cite{su2021benchmarking,tian2017redroid,choudhary2015automated,zeng2016automated,fan2018large,su2020my}. We believe \tool primarily contributes to improving the existing Android GUI testing efforts in the following two aspects. \textit{(1)} The indirect launch strategy for activities proposed by \tool could help the existing tools no longer rely solely on the correctness of the constructed context for activity launching, especially for activities that fail to be launched with the current context information.
    It facilitates the existing testing tools to launch more activities and may finally achieve improvement in the coverage criterion (e.g., activity/method/code coverage). \textit{(2)} Existing Android GUI test suites usually apply random or modeled strategies. The success of AFL \cite{bohme2016coverage} in the binary domain has shown that coverage-based evolutionary algorithms have great potential. Note that activity-based coverage metrics are too coarse from some specific perspectives, for example, there are many scenes that are bound to a single activity, covering the activity does not mean covering all the functionalities in the activity. The fine-grained UI model generated by \tool is helpful in building a scene-based coverage metric. In that case, this more refined metric may motivate the usage and improvement of evolutionary algorithms in Android GUI testing.
    
    \subsection{Limitations}
    Limitations of \tool come from two aspects.
    \textit{(1)} Failure in launching some activities. Despite our proposed smart exploration strategy, some activities still fail to be launched for various reasons, such as the presence of some activities that require authentication (e.g., login), inconsistent activity declarations between the AndroidManifest.xml file and the implementation code, and limited interaction types. We consider \tool could be improved by upgrading the types of components that can be interacted with and by injecting some random system-level events. For indirect launching failure, which may be due to the change of component information during testing, we can design a more reasonable way to record the path of indirect launching for \tool. 
    \textit{(2)} Poor support for non-Native apps. Currently, \tool and most Android GUI testing tools are still for Android native apps \cite{ma2017tale,selvarajah2013native,lee2016hybridroid}; however, HTML5 technology \cite{xanthopoulos2013comparative} and cross-platform development framework have become mainstream in industry \cite{heitkotter2012evaluating, xanthopoulos2013comparative,martinez2017towards}, such as React Native \cite{ReactNative}, Weex \cite{weex}, Kotlin Native\cite{kotlinlang}, Flutter \cite{payne2019developing}, etc., among which Flutter is a cross-platform mobile UI framework strongly supported by Google. 
    In the future, we could work on improving \tool's support for non-native apps.

\section{Related Work}

\subsection{GUI exploration}
GUI exploration is an important way of app abstraction and GUI modeling~\cite{azim2013targeted,yang2018static,chen2019storydroid,lai2019goal,chen2022automatically}. In general, existing work can be divided into two categories according to different goals of GUI exploration.  

\subsubsection{\textbf{GUI exploration for UI modeling}} 
As Android apps are event-driven and composed of activities for user interaction, Activity Transition Graph (ATG)~\cite{azim2013targeted} or Window Transition Graph (WTG)~\cite{yang2018static} is typically used to model the user interface for Android apps. Note that, the extraction has been investigated by both static and dynamic methods.
For example, Yang et al.~\cite{yang2018static} proposed Gator for extracting WTG based on the stack of currently-active windows. The results include the possible GUI window sequences and their associated events and callbacks. Chen et al.~\cite{chen2019storydroid} introduced StoryDroid for statically generating storyboards for Android apps by extracting ATGs along with statically rendered UI pages. StoryDroid combines the results provided by IC3~\cite{octeau2015composite} and ATGs extracted with Fragment and inner class features. 

The most related works are GoalExplorer~\cite{lai2019goal} and StoryDistiller~\cite{chen2022automatically}. Specifically, Lai et al.~\cite{lai2019goal} proposed GoalExplorer, which statically models the UI screens and their transitions between these screens. Apart from the original ATG and WTG, GoalExplorer further extends the static model by adding fragments, drawers, service, and broadcast receivers. Different from this tool, we handle more features of the UI screen through a smart dynamic exploration instead of a static method. StoryDistiller~\cite{chen2022automatically} is an extension of StoryDroid~\cite{chen2019storydroid}, which optimizes the original tool on ATG construction and UI page rendering by combining the original static method and novel dynamic exploration. The strategy of their dynamic exploration is to traverse all clickable components of each UI page that can be launched directly. The goal of dynamic exploration is to obtain new activity transitions that are not parsed in the static method. Compared with StoryDistiller, \tool aims to explore more scenes and scene transitions to construct SceneTG by handling more features such as fragment, drawer, menu, and dialog instead of ATG construction, which is more fine-grained for app UI modeling.

\subsubsection{\textbf{GUI exploration for app testing}}
In the past decade, Android app GUI testing approaches have evolved rapidly, and many testing tools such as Monkey~\cite{developers2012ui}, 
Dynodroid~\cite{machiry2013dynodroid}, Ripper~\cite{amalfitano2012toolset}, A3E~\cite{azim2013targeted}, Sapienz~\cite{mao2016sapienz}, Droidbot~\cite{li2017droidbot}, Stoat~\cite{DBLP:conf/sigsoft/SuMCWYYPLS17}, APEChecker~\cite{fan2018efficiently}, Ape~\cite{gu2019practical}, Humanoid~\cite{li2019humanoid}, Fax~\cite{yan2020multiple}, and PSDroid~\cite{yang2023compatibility}
have been proposed to explore apps and detect bugs or security bugs (\cite{chen2018mobile,chen2020empirical,chen2022ausera}). Since the goal of these tools is to detect more bugs when dynamically testing the apps, the UI transitions are usually incomplete due to the limitation of low activity coverage and test case generation~\cite{chen2019storydroid,chen2022automatically}. 

There are two strategies used in app testing that are related to our work. 
On the one hand, some of them first generated the ATG statically and then conducted dynamic testing based on it. For example, A3E~\cite{azim2013targeted} constructed the ATG by static analysis and leveraged it to guide the dynamic test input generation for app testing. However, many existing works unveiled the statically constructed ATG neglects many activity transitions due to the limitations of static program analysis techniques~\cite{chen2019storydroid,yan2022iccbot}. 
On the other hand, some work focused on dynamic exploration for app testing and after testing, they also provided the UI transition based on the dynamic exploration. For example, Li et al.~\cite{li2017droidbot} proposed DroidBot, a lightweight UI-guided Android test input generator. Apart from the testing results such as test input and identified bugs, DroidBot also generates ATGs for users. Pure dynamic testing has limited activity coverage, significantly restricting ATG completeness. Moreover, the adopted content-based comparison method could produce redundant and duplicate states.

\subsection{ICC resolution}
Researchers have proposed a large number of tools for ICC resolution such as Epicc~\cite{octeau2013effective}, IC3~\cite{octeau2015composite}, IC3DIALDroid\cite{bosu2017collusive},  RAICC\cite{samhi2021raicc}, ICCBot\cite{yan2022iccbot}.
Many works that apply the ICC results have been exhibited for various purposes. In fact, the ICC results also can be used to improve the capability of UI modeling. Yan et al.~\cite{yan2022iccbot} conducted a comprehensive study to evaluate the ICC resolution techniques. According to the results in this paper, we choose ICCBot as a comparison subject to demonstrate the effectiveness of \tool. Compared with the existing ICC resolution, (1) \tool can generate a more complete ATG and SceneTG through both static and dynamic methods. (2) The corresponding UI page of each scene is also provided for users instead of only a graph structure of the UI transitions.

\section{Conclusion}
    In this paper, we proposed \tool, which extracts GUI scenes dynamically by combining three strategies. We present the GUI scenes as a scene transition graph (SceneTG) to model the GUI of Android apps with high transition coverage and fine-grained granularity. Our empirical evaluation has proved the effectiveness and usefulness of \tool. The constructed high-precision model can effectively identify UI updates between different app versions and facilitate developers to design automated regression testing tools and help develop future UI fuzzing testing tools, providing them with effective coverage information.

    
\section*{Acknowledgments}
This work was partially supported by the National Natural Science Foundation of China (Grant No. 62102197, 62102284) and the Natural Science Foundation of Tianjin (Grant No. 22JCYBJC01010).

\balance
\bibliography{references} 

\begin{thebibliography}{10}

\bibitem{chen2019storydroid}
S.~Chen, L.~Fan, C.~Chen, et~al.
\newblock Storydroid: Automated generation of storyboard for android apps.
\newblock In {\em 2019 IEEE/ACM 41st International Conference on Software
  Engineering (ICSE)}, pages 596--607. IEEE, 2019.

\bibitem{azim2013targeted}
T.~Azim and I.~Neamtiu.
\newblock Targeted and depth-first exploration for systematic testing of
  {Android} apps.
\newblock In {\em Proceedings of the 2013 ACM SIGPLAN international conference
  on Object oriented programming systems languages \& applications}, pages
  641--660, 2013.

\bibitem{yang2018static}
S.~Yang, H.~Wu, H.~Zhang, et~al.
\newblock Static window transition graphs for android.
\newblock {\em Automated Software Engineering}, 25:833--873, 2018.

\bibitem{lai2019goal}
D.~Lai and J.~Rubin.
\newblock Goal-driven exploration for android applications.
\newblock In {\em 2019 34th IEEE/ACM International Conference on Automated
  Software Engineering (ASE)}, pages 115--127. IEEE, 2019.

\bibitem{chen2022automatically}
S.~Chen, L.~Fan, C.~Chen, et~al.
\newblock Automatically distilling storyboard with rich features for android
  apps.
\newblock {\em IEEE Transactions on Software Engineering}, 2022.

\bibitem{zhang2023web}
Yuxin Zhang, Sen Chen, and Lingling Fan.
\newblock A web-based tool for using storyboard of {Android} apps.
\newblock In {\em 2023 IEEE/ACM 45th International Conference on Software
  Engineering: Companion Proceedings (ICSE-Companion)}, pages 117--121. IEEE,
  2023.

\bibitem{fan2018efficiently}
Lingling Fan, Ting Su, Sen Chen, Guozhu Meng, Yang Liu, Lihua Xu, and Geguang
  Pu.
\newblock Efficiently manifesting asynchronous programming errors in {Android}
  apps.
\newblock In {\em Proceedings of the 33rd ACM/IEEE International Conference on
  Automated Software Engineering}, pages 486--497, 2018.

\bibitem{yan2022comprehensive}
J.~Yan, S.~Zhang, Y.~Liu, et~al.
\newblock A comprehensive evaluation of android icc resolution techniques.
\newblock In {\em 37th IEEE/ACM International Conference on Automated Software
  Engineering}, pages 1--13, 2022.

\bibitem{su2021benchmarking}
T.~Su, J.~Wang, and Z.~Su.
\newblock Benchmarking automated gui testing for android against real-world
  bugs.
\newblock In {\em Proceedings of the 29th ACM Joint Meeting on European
  Software Engineering Conference and Symposium on the Foundations of Software
  Engineering}, pages 119--130, 2021.

\bibitem{choudhary2015automated}
S.~R. Choudhary, A.~Gorla, and A.~Orso.
\newblock Automated test input generation for android: Are we there yet?(e).
\newblock In {\em 2015 30th IEEE/ACM International Conference on Automated
  Software Engineering (ASE)}, pages 429--440, 2015.

\bibitem{zeng2016automated}
X.~Zeng, D.~Li, W.~Zheng, et~al.
\newblock Automated test input generation for android: Are we really there yet
  in an industrial case?
\newblock In {\em Proceedings of the 2016 24th ACM SIGSOFT International
  Symposium on Foundations of Software Engineering}, pages 987--992, 2016.

\bibitem{li2017data}
Yongfeng Li, Jinbing Ouyang, Bing Mao, Kai Ma, and Shanqing Guo.
\newblock Data flow analysis on android platform with fragment lifecycle
  modeling and callbacks.
\newblock {\em EAI Endorsed Transactions on Security and Safety}, 4(11), 2017.

\bibitem{xiao2019iconintent}
Xusheng Xiao, Xiaoyin Wang, Zhihao Cao, Hanlin Wang, and Peng Gao.
\newblock Iconintent: automatic identification of sensitive ui widgets based on
  icon classification for android apps.
\newblock In {\em 2019 IEEE/ACM 41st International Conference on Software
  Engineering (ICSE)}, pages 257--268. IEEE, 2019.

\bibitem{yan2020multiple}
J.~Yan, H.~Liu, L.~Pan, et~al.
\newblock Multiple-entry testing of android applications by constructing
  activity launching contexts.
\newblock In {\em Proceedings of the ACM/IEEE 42nd International Conference on
  Software Engineering}, pages 457--468, 2020.

\bibitem{yan2022iccbot}
J.~Yan, S.~Zhang, Y.~Liu, et~al.
\newblock Iccbot: fragment-aware and context-sensitive icc resolution for
  android applications.
\newblock In {\em Proceedings of the ACM/IEEE 44th International Conference on
  Software Engineering: Companion Proceedings}, pages 105--109, 2022.

\bibitem{rivest1992md5}
R.~Rivest.
\newblock The md5 message-digest algorithm.
\newblock Technical report, 1992.

\bibitem{Widget}
Android widget listview.
\newblock https://developer.android.com/android/widget/\\ListView, 2022.

\bibitem{googleplay}
Google.
\newblock Google play, 2022.
\newblock https://play.google.com/store.

\bibitem{fdroid}
Ciaran Gultnieks.
\newblock F-droid, 2022.
\newblock https://f-droid.org/.

\bibitem{rountev2017gator}
A.~Rountev, D.~Yan, S.~Yang, H.~Wu, Y.~Wang, and H.~Zhang.
\newblock Gator: Program analysis toolkit for android.
\newblock Technical report, 2017.

\bibitem{ReSeSS}
Github goalexplorer.
\newblock \url{https://github.com/resess/GoalExplorer}, 2022.

\bibitem{do2016redroid}
Q.~C.~D. Do, G.~Yang, M.~Che, et~al.
\newblock Redroid: A regression test selection approach for android
  applications.
\newblock In {\em SEKE}, pages 486--491, 2016.

\bibitem{sharma2019qadroid}
A.~Sharma and R.~Nasre.
\newblock Qadroid: regression event selection for android applications.
\newblock In {\em Proceedings of the 28th ACM SIGSOFT International Symposium
  on Software Testing and Analysis}, pages 66--77, 2019.

\bibitem{peng2021cat}
C.~Peng, A.~Rajan, and T.~Cai.
\newblock Cat: Change-focused android gui testing.
\newblock In {\em 2021 IEEE International Conference on Software Maintenance
  and Evolution (ICSME)}, pages 460--470, 2021.

\bibitem{Preston-Werner}
Tom Preston-Werner.
\newblock Semantic versioning 2.0.0.
\newblock \url{https://semver.org/}, 2022.

\bibitem{V2RayNG}
2dust.
\newblock V2rayng, 2022.
\newblock https://github.com/2dust/v2rayNG.

\bibitem{tian2017redroid}
K.~Tian, G.~Tan, D.~D. Yao, et~al.
\newblock Redroid: Prioritizing data flows and sinks for app security
  transformation.
\newblock In {\em Proceedings of the 2017 Workshop on Forming an Ecosystem
  Around Software Transformation}, pages 35--41, 2017.

\bibitem{fan2018large}
Lingling Fan, Ting Su, Sen Chen, Guozhu Meng, Yang Liu, Lihua Xu, Geguang Pu,
  and Zhendong Su.
\newblock Large-scale analysis of framework-specific exceptions in {Android}
  apps.
\newblock In {\em Proceedings of the 40th International Conference on Software
  Engineering}, pages 408--419, 2018.

\bibitem{su2020my}
Ting Su, Lingling Fan, Sen Chen, Yang Liu, Lihua Xu, Geguang Pu, and Zhendong
  Su.
\newblock Why my app crashes? understanding and benchmarking framework-specific
  exceptions of {Android} apps.
\newblock {\em IEEE Transactions on Software Engineering}, 48(4):1115--1137,
  2020.

\bibitem{bohme2016coverage}
M.~Böhme, V.~T. Pham, and A.~Roychoudhury.
\newblock Coverage-based greybox fuzzing as markov chain.
\newblock In {\em Proceedings of the 2016 ACM SIGSAC Conference on Computer and
  Communications Security}, pages 1032--1043, 2016.

\bibitem{ma2017tale}
Y.~Ma, X.~Liu, Y.~Liu, et~al.
\newblock A tale of two fashions: An empirical study on the performance of
  native apps and web apps on android.
\newblock {\em IEEE Transactions on Mobile Computing}, 17(5):990--1003, 2017.

\bibitem{selvarajah2013native}
K.~Selvarajah, M.~P. Craven, A.~Massey, et~al.
\newblock Native apps versus web apps: Which is best for healthcare
  applications?
\newblock In {\em Human-Computer Interaction. Applications and Services: 15th
  International Conference, HCI International 2013, Las Vegas, NV, USA, July
  21-26, 2013, Proceedings, Part II 15}, pages 189--196. Springer Berlin
  Heidelberg, 2013.

\bibitem{lee2016hybridroid}
S.~Lee, J.~Dolby, and S.~Ryu.
\newblock Hybridroid: static analysis framework for android hybrid
  applications.
\newblock In {\em Proceedings of the 31st IEEE/ACM international conference on
  automated software engineering}, pages 250--261, 2016.

\bibitem{xanthopoulos2013comparative}
S.~Xanthopoulos and S.~Xinogalos.
\newblock A comparative analysis of cross-platform development approaches for
  mobile applications.
\newblock In {\em Proceedings of the 6th Balkan Conference in Informatics},
  pages 213--220, 2013.

\bibitem{heitkotter2012evaluating}
H.~Heitkötter, S.~Hanschke, and T.~A. Majchrzak.
\newblock Evaluating cross-platform development approaches for mobile
  applications.
\newblock In {\em Web Information Systems and Technologies: 8th International
  Conference, WEBIST 2012, Porto, Portugal, April 18-21, 2012, Revised Selected
  Papers 8}, pages 120--138. Springer Berlin Heidelberg, 2013.

\bibitem{martinez2017towards}
M.~Martinez and S.~Lecomte.
\newblock Towards the quality improvement of cross-platform mobile
  applications.
\newblock In {\em 2017 IEEE/ACM 4th International Conference on Mobile Software
  Engineering and Systems (MOBILESoft)}, pages 184--188, 2017.

\bibitem{ReactNative}
React native.
\newblock \url{https://reactnative.dev/}, 2020.

\bibitem{weex}
alibaba.
\newblock weex.
\newblock \url{https://github.com/alibaba/weex}, 2022.

\bibitem{kotlinlang}
Kotlin native.
\newblock \url{https://kotlinlang.org/docs/native-overview.html}, 2022.

\bibitem{payne2019developing}
R.~Payne and R.~Payne.
\newblock Developing in flutter.
\newblock {\em Beginning App Development with Flutter: Create Cross-Platform
  Mobile Apps}, pages 9--27, 2019.

\bibitem{octeau2015composite}
D.~Octeau, D.~Luchaup, M.~Dering, et~al.
\newblock Composite constant propagation: Application to android
  inter-component communication analysis.
\newblock In {\em 2015 IEEE/ACM 37th IEEE International Conference on Software
  Engineering}, volume~1, pages 77--88. IEEE, 2015.

\bibitem{developers2012ui}
A.~Developers.
\newblock Ui/application exerciser monkey, 2012.
\newblock https://developer.android.com/studio/test/monkey.html.

\bibitem{machiry2013dynodroid}
A.~Machiry, R.~Tahiliani, and M.~Naik.
\newblock Dynodroid: An input generation system for android apps.
\newblock In {\em Proceedings of the 2013 9th Joint Meeting on Foundations of
  Software Engineering}, pages 224--234, 2013.

\bibitem{amalfitano2012toolset}
D.~Amalfitano, A.R. Fasolino, P.~Tramontana, et~al.
\newblock A toolset for gui testing of android applications.
\newblock In {\em 2012 28th IEEE International Conference on Software
  Maintenance (ICSM)}, pages 650--653. IEEE, 2012.

\bibitem{mao2016sapienz}
K.~Mao, M.~Harman, and Y.~Jia.
\newblock Sapienz: Multi-objective automated testing for android applications.
\newblock In {\em Proceedings of the 25th International Symposium on Software
  Testing and Analysis}, pages 94--105, 2016.

\bibitem{li2017droidbot}
Y.~Li, Z.~Yang, Y.~Guo, et~al.
\newblock Droidbot: a lightweight ui-guided test input generator for android.
\newblock In {\em 2017 IEEE/ACM 39th International Conference on Software
  Engineering Companion (ICSE-C)}, pages 23--26. IEEE, 2017.

\bibitem{DBLP:conf/sigsoft/SuMCWYYPLS17}
T.~Su, G.~Meng, Y.~Chen, et~al.
\newblock Guided, stochastic model-based gui testing of android apps.
\newblock In {\em Proceedings of the 2017 11th Joint Meeting on Foundations of
  Software Engineering}, pages 245--256, 2017.

\bibitem{gu2019practical}
T.~Gu, C.~Sun, X.~Ma, et~al.
\newblock Practical gui testing of android applications via model abstraction
  and refinement.
\newblock In {\em 2019 IEEE/ACM 41st International Conference on Software
  Engineering (ICSE)}, pages 269--280. IEEE, 2019.

\bibitem{li2019humanoid}
Y.~Li, Z.~Yang, Y.~Guo, et~al.
\newblock Humanoid: A deep learning-based approach to automated black-box
  android app testing.
\newblock In {\em 2019 34th IEEE/ACM International Conference on Automated
  Software Engineering (ASE)}, pages 1070--1073. IEEE, 2019.

\bibitem{yang2023compatibility}
Sen Yang, Sen Chen, Lingling Fan, Sihan Xu, Zhanwei Hui, and Song Huang.
\newblock Compatibility issue detection for {Android} apps based on
  path-sensitive semantic analysis.
\newblock In {\em 2023 IEEE/ACM 45th International Conference on Software
  Engineering (ICSE)}, pages 257--269. IEEE, 2023.

\bibitem{chen2018mobile}
Sen Chen, Ting Su, Lingling Fan, Guozhu Meng, Minhui Xue, Yang Liu, and Lihua
  Xu.
\newblock Are mobile banking apps secure? what can be improved?
\newblock In {\em Proceedings of the 2018 26th ACM Joint Meeting on European
  Software Engineering Conference and Symposium on the Foundations of Software
  Engineering}, pages 797--802, 2018.

\bibitem{chen2020empirical}
Sen Chen, Lingling Fan, Guozhu Meng, Ting Su, Minhui Xue, Yinxing Xue, Yang
  Liu, and Lihua Xu.
\newblock An empirical assessment of security risks of global {Android} banking
  apps.
\newblock In {\em Proceedings of the ACM/IEEE 42nd International Conference on
  Software Engineering}, pages 1310--1322, 2020.

\bibitem{chen2022ausera}
Sen Chen, Yuxin Zhang, Lingling Fan, Jiaming Li, and Yang Liu.
\newblock Ausera: Automated security vulnerability detection for {Android}
  apps.
\newblock In {\em Proceedings of the 37th IEEE/ACM International Conference on
  Automated Software Engineering}, pages 1--5, 2022.

\bibitem{octeau2013effective}
D.~Octeau, P.~McDaniel, S.~Jha, et~al.
\newblock Effective inter-component communication mapping in android with
  epicc: An essential step towards holistic security analysis.
\newblock {\em Effective Inter-Component Communication Mapping in Android with
  Epicc: An Essential Step Towards Holistic Security Analysis}, 2013.

\bibitem{bosu2017collusive}
A.~Bosu, F.~Liu, D.~Yao, et~al.
\newblock Collusive data leak and more: Large-scale threat analysis of
  inter-app communications.
\newblock In {\em Proceedings of the 2017 ACM on Asia Conference on Computer
  and Communications Security}, pages 71--85, 2017.

\bibitem{samhi2021raicc}
J.~Samhi, A.~Bartel, T.F. Bissyandé, et~al.
\newblock Raicc: Revealing atypical inter-component communication in android
  apps.
\newblock In {\em 2021 IEEE/ACM 43rd International Conference on Software
  Engineering (ICSE)}, pages 1398--1409. IEEE, 2021.

\end{thebibliography}

\end{document}